\documentstyle[epsfig]{mn}
\begin{document}
\input BoxedEPS.tex
\newcommand{\aip}{{\small ${\cal AIPS}$}}
\newcommand{\gtsim}{\mbox{{\raisebox{-0.4ex}{$\stackrel{>}{{\scriptstyle\sim}}
$}}}}
\newcommand{\ltsim}{\mbox{{\raisebox{-0.4ex}{$\stackrel{<}{{\scriptstyle\sim}}
$}}}}
\newcommand{\s}{$\stackrel{\rm s}{.}$}
\newcommand{\h}{$^{\rm h}$}
\newcommand{\m}{$^{\rm m}$}
\newcommand{\pp}{$\stackrel{\prime\prime}{.}$}
\newcommand{\de}{$^{\circ}$}
\newcommand{\p}{$^{\prime}$}
\newcommand{\arc}{$^{\prime\prime}$}
\newcommand{\marc}{^{\prime\prime}}
\newcommand{\rs}{{\em $r_s$}}
\newcommand{\DPM}{{\em DPM}}
\newcommand{\alf}{{\displaystyle\biggl({\nu_{\rm h} \over \nu_{\rm l}}\biggr)^{\alpha}} }

\newcommand{\figstart}[1]
    { \begin{figure}[htb]
      \begin{picture}(0,#1) }
\newcommand{\figend}[4]
    { \end{picture}
      \special{#1}
      \caption[#2]{#3}
      \label{#4}
      \end{figure} }
\newcommand{\fig}[5]
    { \figstart{#1}
      \figend{#2}{#3}{#4}{#5} }
\newcommand{\bHS}{\beta_{\mbox{\scriptsize HS}}}
\newcommand{\bBF}{\beta_{\mbox{\scriptsize BF}}}
\newcommand{\nT}{\nu_{\mbox{\scriptsize T}}}
\newcommand{\et}{E_{\mbox{\scriptsize T}}}
\newcommand{\nTn}{\nu_{\mbox{\scriptsize Tn}}}
\newcommand{\nTf}{\nu_{\mbox{\scriptsize Tf}}}
\newcommand{\tn}{\tau_{x\mbox{\scriptsize n}}}
\newcommand{\tf}{\tau_{x\mbox{\scriptsize f}}}
\newcommand{\xn}{x_{\mbox{\scriptsize n}}}
\newcommand{\xf}{x_{\mbox{\scriptsize f}}}
\newcommand{\yn}{y_{\mbox{\scriptsize n}}}
\newcommand{\yf}{y_{\mbox{\scriptsize f}}}
\newcommand{\lln}{l_{\mbox{\scriptsize n}}}
\newcommand{\llf}{l_{\mbox{\scriptsize f}}}
\newcommand{\Dn}{f(\Delta_{\mbox{\scriptsize n}})}
\newcommand{\Df}{f(\Delta_{\mbox{\scriptsize f}})}
\newcommand{\B}{\mbox{$B$}}
\newcommand{\Bo}{\mbox{$B$}_{0}}

\SetRokickiEPSFSpecial
\HideDisplacementBoxes


\title[Massive Star Formation in Galaxies]{Massive Star Formation in Galaxies:
 Radiative transfer models of the UV to mm emission of starburst galaxies.}
\author[Efstathiou,  Rowan-Robinson and  Siebenmorgen]
{A.Efstathiou$^1$, M.Rowan-Robinson$^1$ and R.Siebenmorgen$^2$\\
$^1$Astrophysics Group, Imperial College London, Blackett Laboratory,
Prince Consort Road, London SW7 2BZ;\\ 
$^2$ISO Science Operations Centre, Astrophysics Division of ESA,
   Villafranca del Castillo, P.O.Box 50727, E-28080 Madrid 
}
\maketitle
\begin{abstract}

 We present illustrative models for the UV to millimeter emission of
starburst galaxies which are treated as an ensemble of optically thick
giant molecular clouds (GMCs) centrally illuminated by recently formed
stars. The models follow the evolution of the GMCs due to the
ionization-induced expansion of the HII regions and the evolution of
the stellar population within the GMC according to the Bruzual \&
Charlot stellar population synthesis models.  The effect of
transiently heated dust grains/PAHs to the radiative transfer, as well
as multiple scattering, is taken into account.

 The expansion of the HII regions and the formation of a narrow
neutral shell naturally explains why the emission from PAHs dominates
over that from hot dust in the near to mid-IR, an emerging characteristic
of the infrared spectra of starburst galaxies.

 The models allow us to relate the observed properties of a galaxy to
its age and star formation history. We find that exponentially
decaying $10^7-10^8$ yrs old bursts can explain the IRAS colours of
starburst galaxies.  The models are also shown to account
satisfactorily for the multiwavelength data on the prototypical
starburst galaxy M82 and NGC6090, a starburst galaxy recently observed
by ISO. In M82 we find evidence for two bursts separated by
$10^7$yrs. In NGC6090 we find that at least part of the far-IR excess
may be due to the age of the burst ( $6.4 \times 10^7$yrs). We also
make predictions about the evolution of the luminosity of starbursts
at different wavelengths which indicate that far-IR surveys may
preferentially detect older starbursts than mid-IR surveys.

\end{abstract}


\section{Introduction} 

 It now looks increasingly clear that the long-sought era of galaxy
formation is becoming accessible with a number of observational
methods for close scrutiny. These range from optical/ultraviolet
studies (Steidel \& Hamilton 1992, Lilly et al 1996, Madau et al
1996), mid-infrared (Rowan-Robinson et al 1997 and references therein)
and sub-millimeter surveys (Hughes et al 1998). A number of authors
have discussed the history of star formation in the Universe which
appears to have peaked at a redshift of about 1-3. Estimates like this
are, of course, model-dependent as they involve the converion from an
observed (usually monochromatic) luminosity density to a
star-formation rate. In particular, they depend on the extent and
geometry of dust obscuration especially at the shorter wavelengths.

 As is well known a large fraction of the power emitted by galaxies
(ranging from about 30\% in normal galaxies to almost 100\% in
actively star forming galaxies or starbursts) lies in the infrared
part of the spectrum as a result of reprocessing of starlight by
dust. Extensive infrared observations of galaxies are therefore
necessary in order to describe fully their energy output. With the
advent of the Infrared Space Observatory (ISO) the spectra of a number
of galaxies in the local Universe have been observed with
unprecedented detail in the infrared. They invariably display a
variety of absorption/emission features due to dust/molecules. It is
clear that radiative transfer calculations in dusty media will be
useful for the interpretation of these observations and a development
of a better understanding of the origin of the infrared luminosity of
galaxies.

 Radiative transfer models for the infrared emission of starburst
galaxies have been presented before by Rowan-Robinson \& Crawford
(1989; hereafter RRC), Rowan-Robinson \& Efstathiou (1993; hereafter
RRE), Kr\"ugel \& Siebenmorgen (1994; hereafter KS). These models used
state of the art codes for calculating the transfer of radiation in
dusty media and incorporated a model for the composition and size
distribution of grains in the interstellar medium. KS additionally
included the effect of transiently heated grains/PAHs and considered
the local change of dust temperatures in hot spots around luminous OB
stars.  The basic assumption of previous starburst models is that a
starburst is made up of an ensemble of compact HII regions similar to
those found in our galaxy. A number of authors (e.g. Rowan-Robinson
1980, Kr\"ugel \& Walmsley 1984, Churchwell, Wolfire \& Wood 1990,
Efstathiou \& Rowan-Robinson 1994) studied the infrared properties of
different samples of such HII regions and concluded that they could be
modelled adequately by spherically symmetric dust clouds surrounding
young massive stars.

All of the above models assume that the cloud  ensemble in the
starburst consists of a number of  identical systems.
 In this paper we present illustrative models for the evolution of
giant molecular clouds (GMCs) induced by massive star formation at their
centers and calculate their infrared spectra. In the radiative
transfer code we use, the effect of transiently heated
particles/PAHs as well as classical grains is included. We also follow the
evolution of the stellar populations with the models of Bruzual \&
Charlot (1995). This approach allows us to relate the observed
properties of a starburst to its age and its star formation history.
We illustrate these models by comparison with the IRAS colours of starburst
galaxies as a class and with the multiwavelength data on M82 and NGC6090.

\section{A new starburst model}

   The basic assumption of our model for a starburst galaxy is that star
formation takes place primarily within optically thick molecular
clouds. This is supported by an array of observational studies
(e.g. Elmegreen 1985) which show that molecular clouds are associated
with  young stars. 

   The mass of molecular clouds in our Galaxy ranges between
 $10^2-10^7M_\odot$ with a mass distribution approximately following
 $M ^{-1.5\pm0.1}$ (Dame et al 1983, Solomon et al 1987).  It follows
 that about $70\%$ of molecular mass is associated with GMCs more
 massive than $10^6 M_\odot$. If the mass distribution in a starburst
 follows a similar form as the Galactic one, and GMCs are at least as
 likely to be the sites of massive stars as the less massive molecular
 clouds, then we would expect the bulk of the luminosity of the
 starburst to arise from GMCs with a fairly narrow range of mass. In
 fact low-mass molecular clouds, like the nearby Taurus and Ophiuchus
 clouds, are known to form predominantly low-mass stars.

  An indirect argument that massive star formation in galaxies takes
  place in GMCs  with roughly a power-law mass distribution comes from
  HST observations of starburst galaxies (O'Connell et al 1995, Meurer
  et al 1995) which  reveal a population of super star clusters
  following a luminosity function of a power-law form $\phi(L)
  \propto L^{-2}$. A similar slope is found for systems of young
  clusters in other galaxies (e.g. Whitmore \& Schweizer 1995). 
  This luminosity function is quite unlike that of
  globular clusters although it has been suggested that it could
  evolve into one (Meurer et al). In our model these star clusters
  represent the evolved counterparts of  star clusters forming within
  GMCs, with the mass spectrum given above, after they have dispersed
  their nascent  molecular clouds. It is not clear what the ages of
  these clusters are because of the known age-reddening degeneracy but
  in the case of M82 their proximity to the 2$\mu m$ nucleus (see
  section 4) points towards an age $> $ 10Myrs.

 In previous studies we
assumed that a starburst is an ensemble of identical star forming
complexes, which we approximate as spherical. Here we refine this
model by considering the fact that, given the starburst takes place
over a finite period of time, the star-forming complexes that
constitute it are bound to be at different evolutionary stages. We
therefore use a simple evolutionary model for HII regions to construct
a family of models that predict the infrared spectrum of HII regions
as a function of their age. 

\subsection{The dusty  HII region phase}

Star formation takes place primarily in the dense cores of GMCs. The
details of the physical processes involved are not well understood
yet. The efficiency of star formation (or the gas consumption rate)
ranges from about $ 1\%$ in late type spirals to $60\%$ or more in
starburst galaxies (Kennicutt 1998).

 Once massive stars form they can ionize the surrounding medium and
  inhibit further star formation. This in itself, however, cannot
  explain the low star formation efficiency in disk galaxies
  because it does not explain what determines the number of massive
  young stars formed initially.

The evolution of GMCs is determined by ionization induced expansion in
the early stages ($t < 10^7$ yrs) and later by stellar winds and
supernova explosions. The latter eventually disperse the molecular
clouds on timescales of a few times $10^7$ years.  Current scenaria of
self-propagating star formation (e.g. Larson 1988, Tenorio-Tagle \&
Bodenheimer 1988, hereafter TTB) hold that molecular cloud formation
by a number of mechanisms occurs on a timescale of $10^8$ yrs, to
complete the cycle.

The evolution of HII regions due to ionization has been the subject of
extensive study both analytically and numerically for a broad variety
of circumstances (spherically symmetric case, two-dimensional solutions
leading to so-called champagne flows) and is considered to be
generally well understood (for a review see Franco et al 1990, TTB).

The presence of a number of massive stars in the centre of a GMC
producing numerous ionizing photons leads to the formation, on an
extremely short timescale, of the classical HII region with initial
Stromgren radius $R_S$. The pressure in the ionized region, being
orders of magnitudes higher than in the rest of the GMC, drives the
expansion of the HII region into the surrounding medium. The details
of the expansion depend on a number of factors including the density
distribution in the medium. Assuming a constant density medium and
spherical symmetry, the evolution of the HII region radius can be
described by (Spitzer 1978) 

\begin{equation}
  R= R_S \; \left[ 1 + {7 \over 4} {{c_i t} \over R_S} \right]^{4
\over 7}  \label{eq1}
\end{equation}
where $c_i$ is the sound speed in the ionized gas. $t$ is the time
since the formation of the initial Stromgren sphere. However, since the
timescale for the latter is extremely short we will assume that $t$ is
the time from the onset of star formation (assumed to take place
instantaneously) in the GMC.
 
The expansion can be broken down into a number of phases, marked by
the time when massive stars first move off the main sequence and some
recombination occurs ($t \approx 3-4 \times 10^6$ yrs), a phase of further
ionization, and finally total recombination ($t \approx 10^7$ yrs) as $F_*$
diminishes. However, throughout expansion and until the supernovae
ejecta finally reach the expansion front caused by ionization,
equation (1) can be considered to give a reasonable description of the
evolution, a fact confirmed by numerical simulations (TTB).

In the simple case of a core of constant density $n_c \; cm^{-3} $ the
initial Stromgren radius $R_S$ generated by a compact star cluster producing
$F_*$ ionizing photons is given by (Franco et al 1990),
\begin{equation}
 R_S = 4.9 \; \left({F_{*} \over {5 \times 10^{52} s^{-1}
              } }\right)^{1 \over 3}  
       \; \left({n_c \over {2 \times
        10^3 cm^{-3}}}\right)^{-{2 \over 3}} \; pc \label{eq2}
\end{equation}
	
The average densities of GMCs in our Galaxy are in the range $10$ to
$10^2 cm^{-3}$ (Dame et al 1983) but their cores, where most of the
stars form, have densities three or more orders of magnitude higher.
Higher densities are also deduced in more actively star-forming
galaxies in accordance with the Schmidt law (Kennicutt 1998). Given
the lack of knowledge about $n_c$, its density distribution, the
effect of dust within the HII region and our assumption that all star
formation in a given GMC occurs instantaneously on defining $R_S$,
equation (2) provides only a rough estimate of $R_S$, probably an
upper limit.

We use the tables of Bruzual and Charlot (1995) to derive the number
of Lyman continuum photons $F_*$. For a Salpeter IMF and stars in the
mass range $0.1-125 M_{\odot}$,

\begin{equation}
 F_* = 5 \times 10^{52} \; \left({\eta \over 0.25}\right) \; \left({M_{GMC}
\over 
{10^{7} M_\odot } }\right)  \; s^{-1} \label{eq3}
\end{equation}
where $\eta$ is the star formation efficiency.

The molecular cloud radius $r_2$ is related to its  mass $M_{GMC}$ and
average density $n_{av}$ (initially assumed to be uniform) by

\begin{equation}
 r_2 = 50 \; \left({(1-\eta) M_{GMC} \over {10^{7} M_\odot } }\right)^{1 \over 3}  
       \; \left({n_{av} \over {300 cm^{-3}}}\right)^{-{1 \over 3}} \;
       pc \label{eq4}
\end{equation}

At time $t=0$, the GMC is essentially divided into three zones.
Zone A (inside the dust sublimation radius $r_1$) consists of ionized
gas and the stellar cluster which is approximated as a point source. In
zone B ( $ r_1 < r < R_S$) we have ionized gas and dust, whereas zone
C ( $ R_S < r < r_2 $) is the dusty neutral zone.

As the expansion gets under way, the shock wave at $R(t)$, leading the
ionization front at $R_i(t)$,  accumulates neutral gas between $R_i$
and $R$. Inside $R_i$ the mass density of the ions $\rho_i$ follows
(Franco et al )

\begin{equation}
\rho_i (t) = \rho_{n_0}  \; \left({R \over {R_S}}\right)^{-3/2}
\label{eq5}
\end{equation}
where $\rho_{n_0}$ is the initial mass density of the neutral material
assumed to be equal to the initial mass density of the ions.
The density inside the HII region is probably constant to a good
approximation as the expansion is subsonic (Franco et al ).

The density enhancement and the separation of $R_i$ and $R$ cannot be
followed analytically. Instead we assume that the neutral gas
accumulated in the shock is spread over the entire neutral cloud, and
we use the principle of conservation of mass to calculate the
(uniform) density of neutral material $\rho_n (t)$

\begin{equation}
 \rho_n (t) = \rho_{n_0} {{ 1 - \left({{R} \over {r_2}}\right)^{3/2}
                \; \left({{R_S} \over {r_2}}\right)^{3/2}}
                 \over {1 - \left({{R} \over {r_2}}\right)^{3}}}
                 \label{eq6}
\end{equation}                 
      
Assuming the standard conversion from gas column density to extinction
(Savage \& Mathis 1979), the visual extinction to the center of the
cloud at $t=0$ is given by,

\begin{equation}
\tau_{V_0} = 50 \; \zeta \; \left({{r_2} \over {50 pc}}\right) \; 
\left({ n_{av} \over {300 cm^{-3}} }\right)
\label{eq7}
\end{equation}
where $\zeta $ is the metallicity with respect to solar.
      
To estimate the evolution of the optical depth of the cloud we assume
that the number density of grains scales in the same way as $\rho_i$
and $\rho_n$ to get

\begin{equation}
  \tau_{V} =  \tau_{V_0} \; [{\rho_n \over \rho_{n_0}} (1 - {R \over
r_2}) \; + f_d {\rho_i \over \rho_{n_0}} ({R \over r_2} -  {r_1 \over 
r_2}) ]  \label{eq8}
\end{equation}                 
where $f_d$ takes account of a possible depletion of dust in the HII region as
a result of  shocks etc.

We will assume that the shock advances into the neutral part of the
GMC until the swept up material is half of the original GMC mass. For
our assumed uniform density medium this means the advance of the shock
stops when $R \approx 0.8 r_2$

\subsection{The supernova phase}

  The state of the GMC at $\sim 10^7$ years is likely to be a narrow
 neutral shell still expanding, because of the momentum it acquired in
 the HII phase, at about 10 km s$^{-1}$. 
 Over the next $3-4 \times 10^7$ years supernova explosions are
 likely to be the main factor that will influence the evolution of the
 GMCs eventually leading to their dispersal and return of most of the
 gas to the HI phase.

 This phase has also been extensively studied analytically and
 numerically, but unfortunately mainly for the `standard' density case
 of 1 cm$^{-3}$ (TTB and references therein). Most relevant for our
 purposes is the work of McCray and Kafatos (1987) who presented an
 analytical solution for the case of multi-supernova explosions in an
 OB association. Numerical solutions have shown that this solution is
 an adequate approximation if the supernova explosions are frequent as
 in our case.  The sequence of discrete supernova explosions in the
 model is replaced by a scaled-up stellar wind solution. The radius of
 the swept up shell before the hot interior begins to cool at time
 $t_c$ is given by

\begin{equation}
 R_{SN} = 97 \left({{N_*E_{51}} \over {n}}\right)^{0.2} \left({{t} \over
     {10^7yrs}}\right)^{0.6} pc \label{eq9}
\end{equation}                 
       where $N_*$ is the
total number of stars in the association (with $M > 7
M_{\odot}$) that are destined to explode as supernovae , $ E_{51}$ is
the energy per supernova (assumed constant) in units of $10^{51}$
ergs, and $n$ is the density of the medium, again assumed to be constant.

After time $t_c$, given by

\begin{equation}
  t_c =4 \times 10^6 \zeta^{-1.5} (N_* E_{51})^{0.3} n^{-0.7} yr \label{eq10}
\end{equation}                 
the shell expands as
\begin{equation}
  R_{SN} = R_c (t/t_c)^{1/4} \label{eq11}
\end{equation}                       
where $R_c$ is the radius reached at $t_c$.

 For a Salpeter IMF and stars in the mass range $0.1-125 M_{\odot}$
the Bruzal \& Charlot models predict $8.92 \times 10^{-3}$ supernova
explosions per 1 $M_{\odot}$ of stellar mass. So, our adopted
parameters for the GMCs in a starburst ($M_{GMC} = 10^7 M_{\odot}$,
$\eta = 0.25$; see section 3) imply $N_*=2.23
\times 10^4$. This, in turn, implies that  the supernova shell will
remain inside the shell formed by ionization at $t=10^7$yrs  if the
density of the medium within which the supernovae are exploding is
higher than $10^3 cm^{-3}$. This is not unreasonable given the
contributions from stellar winds, failed `cores' etc. in a compact
cluster of this mass. The velocity of the supernova shell is also
predicted to drop well below the expansion velocity of the neutral
shell by $10^7$ yrs.

 In this paper we will therefore assume that the supernova shell
  remains `trapped' inside the neutral shell which  continues to
 expand at 10 km s$^{-1}$ until $t=4 \times 10^7$yrs. 
 We discuss whether this assumption is inconsistent with infrared
 observations of galaxies later in the paper.

 Our plan in this paper is to use the simple evolutionary scenario
outlined in the last two sections and our radiative transfer code in
dusty media to predict the infrared properties of evolving HII
regions. A further ingredient of our models will be the evolutionary
synthesis models of Bruzual \& Charlot (1995) which will provide the
spectral energy distribution of the evolving stellar population
powering the HII region.
 
The family of evolving HII region models will then form the basis of our
starburst models under our assumption that the latter are an ensemble
of giant HII regions at different evolutionary stages.

In order to get a first impression of the characteristics of our
models we have in this paper confined our attention to the spherically
symmetric case. Clearly, highly non-spherical geometries can arise
especially at the latter stages of the evolution and we plan to explore
these situations in  future studies.

\subsection{Dust model}

 The model we have assumed for the absorption/emission properties of
 the dust is an extension of the `classical' grain model (e.g. Mathis
 1990 and references therein) to take into account the effects of small
 grains and molecules. The model is described in detail in
 Siebenmorgen \& Kr\"ugel (1992; hereafter SK92) where it has been
 shown to account satisfactorily for the emitted spectra of dust in a
 number of environments (solar neighbourhood, planetary nebulae,
 star-forming regions).

 The model assumes three populations of grains and aims to fit the
 average interstellar extinction curve subject to the constraints
 imposed by abundances of heavy elements that are in the solid state
 etc.  The first population of grains consists of the large grains
 (assumed to have optical properties as the ``astronomical'' silicates
 of Draine \& Lee 1984 and of amorphous carbon by Edoh 1983). These
 grains provide the bulk of the emission/absorption at long
 wavelengths and are responsible for the silicate resonances at 9.7
 and 18$\mu m$. The large  grains are assumed to have a power-law size
 distribution ($n(a) \propto a^{-q}$, $q=3.5$, 100\AA $\leq a \leq
 $2500\AA).

 The second population of grains consists of small graphite particles 
($n(a) \propto a^{-q}$, $q=4$, 10\AA $\leq a \leq  $100\AA). These
grains are responsible for the 2175\AA ~~bump (Draine 1989) and emit
primarily in mid-IR wavelengths. Because of their small size they show
dramatic fluctuations in temperature. Their emission is calculated
with the method of Siebenmorgen et al (1992) which is a faster (but
still fairly accurate) alternative to the treatment of Guhathakurta
\& Draine (1989).

 The third grain population is composed of PAHs that are now widely
 believed to be the carriers of the infrared features at 3.3, 6.2,
 7.7, 8.6 and 11.3$\mu m$ (Puget \& Leger 1989, Allamandola et al 1989).  
 The SK92 model assumes two components for the PAHs: a single molecule
 component made up of 25 carbon atoms and a cluster component made
 up of 10-20 molecules. A ratio of carbon atoms in PAHs to H atoms in
 the ISM of  $3 \times 10^{-5}$ is assumed (which is about 10\%
 of the total carbon abundance thought to reside in grains/molecules).
 The PAH clusters dominate at longer (mid-IR) wavelengths and are largely
 responsible for the quasi-continuum underlying the features (Desert,
 Boulanger \& Puget 1990). The PAHs absorb mainly in the Far UV and to
 a lesser extent in the visible (SK92).

 The PAHs are thought to be dehydrogenated depending on their
 environment, so a degree of dehydrogenation $\alpha_{H/C}$ (defined
 as the ratio of hydrogen to carbon atoms) needs to be introduced.
 $\alpha_{H/C}$ varies from about 0.4 for the solar neighbourhood to
 0.1 or less for star forming regions (SK92). As the emission of some
 of the features (3.3, 8.6 and 11.3$\mu m$) is proportional to the
 number of hydrogen atoms and of others (6.2 and 7.7$\mu m$)
 proportional to the number of carbon atoms, $\alpha_{H/C}$ can have
 an effect on the relative strengths of the features.

 In regions of high radiation intensity (such as those to be found in
 regions of massive star formation or AGN), the small graphites and
 PAHs are thought to be underabundant with respect to the ISM by
 factors of 10 or more (SK92, Rowan-Robinson 1992 and references
 therein).  This will result in a flattening of the extinction curve
 in the UV and an elimination of the 2175\AA ~feature.  The latter has
  been shown to be a characteristic of the extinction curves
 of starburst galaxies (Gordon, Calzetti \& Witt 1997).  For the
 models presented in this paper we assume that all the grains are
 depleted by a factor of 5 (i.e. $f_d=0.2$) inside $R$. We further
 assume that the clusters are made of 500 C atoms and contain $90\%$
 of the total C abundance in PAHs. Note that the PAHs abundance is further
 reduced because of the photo-destruction mechanism that is
 self-consistently applied in the code (Siebenmorgen 1993).

\subsection{Radiative transfer model}

The method of solution of the radiative transfer problem in dusty
media containing transiently heated grains is described in Efstathiou
\& Siebenmorgen (1999).  The method of obtaining the intensity
distribution at any point in the cloud and hence iterating for the
temperature of each of the large grains is that used by Efstathiou \&
Rowan-Robinson (1990, 1995; hereafter ER90, ER95 respectively). The
emission of the transiently heated particles is calculated according
to the method of Siebenmorgen et al. (1992). Proper treatment of the
photodestruction of the PAHs and the sublimation of the large grains,
at the inner part of the cloud, is taken into account.

\section{Evolving HII region models}

\begin{figure*}
\epsfig{file=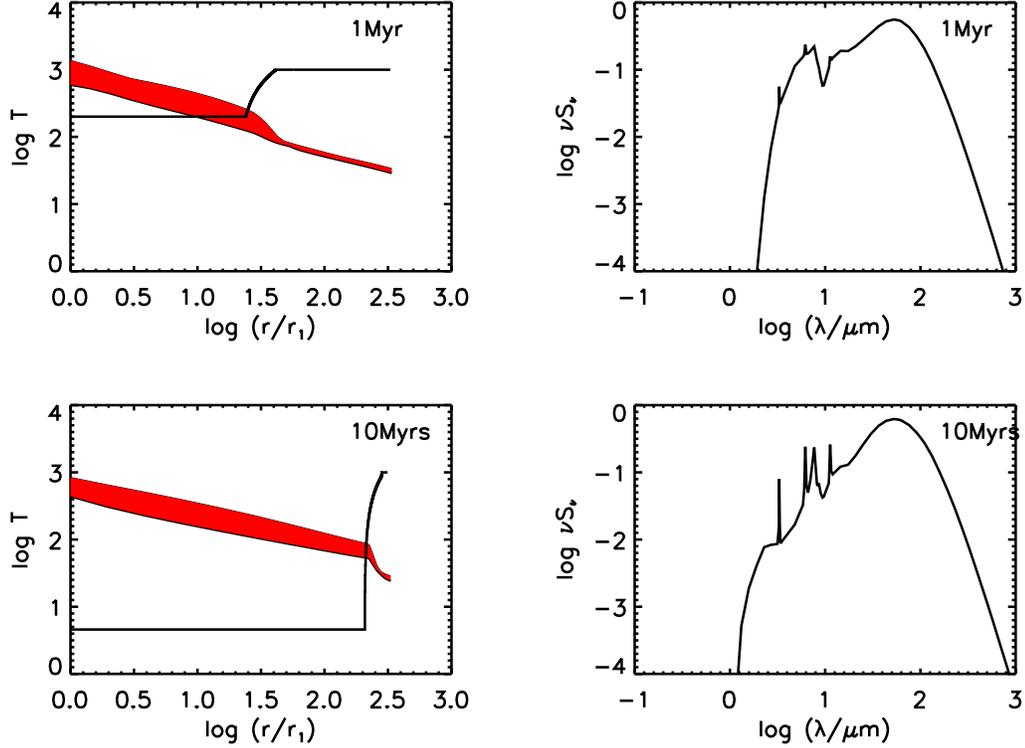, angle=0, width=15cm}
\caption{Temperature and density distributions (left)  and spectral
 energy distributions (right), of the GMCs at two representative
  ages of the HII region phase.  The shaded regions show the spread in
  temperature attained by the different grain species. The density
  and $\nu S_{\nu} $ scale is arbitrary.}\label{hii}
\end{figure*}

 There are basically three  free parameters in our model
 ($M_{GMC}$, $\eta$, and $n_{av}$) which we fix by relating to
 observational constraints where available.

 Kennicutt (1998) finds the median rate of gas consumption in
 starburst galaxies per $10^8$ years to be  $\sim 30\%$. Colbert et
 al (1998) estimate from ISO spectroscopy that the
 mass in stars in the M82 starburst is $0.5-1.3 \times 10^8
 M_\odot$. Assuming the molecular gas mass of $\sim 2 \times 10^8
 M_\odot$ estimated from a number of studies (e.g. Hughes et al 1994)
 is associated with the GMCs that formed those stars, the implied gas
 consumption rate is 0.2-0.39. In this paper we assume $\eta =0.25$.

 For $M_{GMC}$ we choose $10^7 M_\odot$, a mass close to the upper
 limit of the galactic mass spectrum, as the emission is likely to be
 dominated by such GMCs (section 2). As a check, we can compare the
 luminosity expected from one of these GMCs ($M_{GMC}=10^7 M_\odot,
 \eta=0.25$) with that of the 10$\mu m$ knots in M82. Assuming they
 have the same spectrum as the starburst as a whole, each of these
 knots has a luminosity of $\sim 2 \times 10^8 L_\odot$ (Telesco \&
 Gezari 1992). By comparison, the models  of Bruzual \& Charlot ($0.1 < M <
 125M_\odot$) predict $L \sim 3.5 \times 10^8 L_\odot$ for a $10^7$
 years old instantaneous burst.

 Probably the most uncertain free parameter in our models is the
 initial GMC average density $n_{av}$ and the core density $n_c$. We
 assume $n_{av} = 300 cm^{-3}$ which is within the range of GMC
 densities found in the centre of the galaxy (G\"{u}sten 1989). There
 is evidence (e.g. RRE, Downes \& Solomon 1998) that higher gas
 densities are to be found in the more extreme starbursts powering
 ultraluminous infrared galaxies.  The core density is assumed to be
 $2 \times 10^3 cm^{-3}$.

 In Figure 1 we plot the density, temperature and spectral energy
 distributions of a GMC with assumed parameters $M_{GMC}=10^7 M_\odot,
 \eta=0.25$, and $n_{av}=300 cm^{-3}$. These parameters are used for the
 remainder of this paper. The density at the boundary between the
 neutral and ionized region has been smoothed in order to be able to
 handle it numerically. The shading in the same diagram represents the
 spread in temperature between different grain species. Note that the
 spread is much smaller in the neutral region because it is more
 optically thick (see Efstathiou \& Rowan-Robinson 1994, Kr\"ugel \&
 Walmsley 1984 for a discussion of such radiative transfer effects in
 dust clouds). Our evolutionary scheme predicts that by $10^7$ years
 the ionization front has compressed the dust to a very narrow shell
 ($r_1/r_2 \approx 0.8$).

 The SEDs of the GMCs vary significantly with age. In the early stages
  of the cloud's evolution its SED is predicted to be warmer and show
  little signs of PAH emission. This is partly because the stellar
  population is younger and the radiation field stronger. The main
  contributing factor though is that there is more hot dust inside the
  Stromgren sphere than at later times. The weakness of the PAH
  features is partly due to the stronger near to mid-IR continuum
  emission from large grains and to the higher degree of
  photodestruction because of the stronger and harsher radiation
  field.

 By 10Myrs the mid-IR spectrum is dominated by the PAHs features and
 quasi-continuum and shows the characteristic shape that is observed in
 the spectra of starburst galaxies (Willner et al 1977, Acosta-Pulido
 et al 1996). Note that while the
 general trend is for the dust shell to cool off with time, the
 expansion of the HII region introduces a subtle effect that somewhat
 counteracts this in the HII phase. As the neutral shell is pushed out
 and the density inside $R$ declines, the total optical depth to the
 stellar cluster (eqn 8) decreases.  This means that the neutral shell
 actually heats up a little bit. This effect is better demonstrated by
 the evolution of the IRAS colours, plotted in Figure 2.

 At $t > 20Myrs$ the peak of the SEDs of the shells shifts to longer
 wavelengths and become remarkably similar to those of the diffuse
 medium and cirrus clouds.

 The effect of changing the IMF from Salpeter to Miller-Scalo is to
 reduce $F_*$ by about 80$\%$ and $R_S$ by about 20$\%$. So the effect
 on the overall SEDs is small.

\section{Evolving starburst models}

\begin{figure*}
\epsfig{file=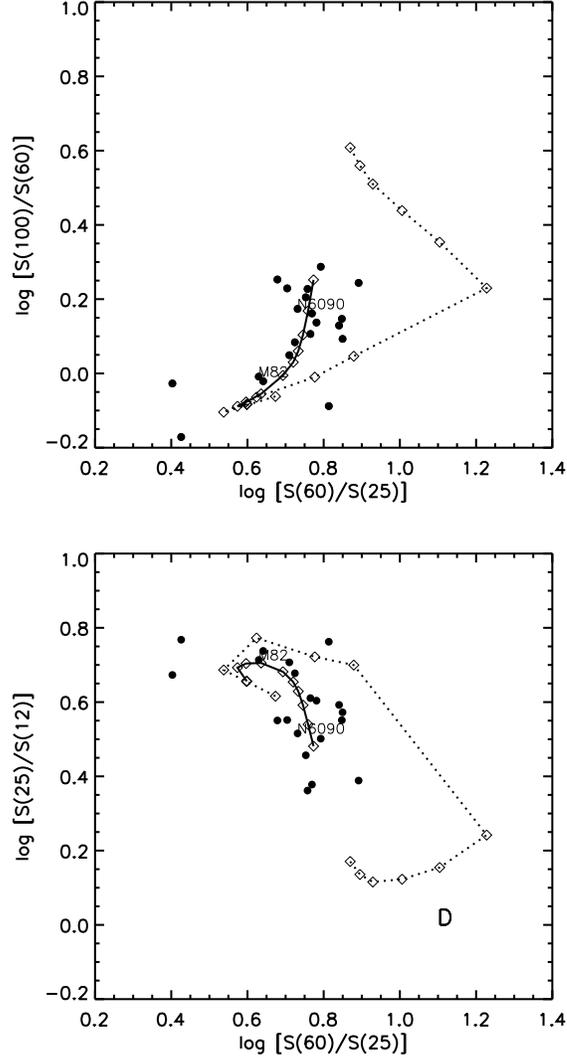, angle=0, width=8cm}
\caption{IRAS colour-colour diagrams of starburst galaxies from the
sample of Rowan-Robinson \& Crawford (1989) with models
superimposed. The dotted line shows the colours of the sequence of
GMCs. The ages indicated by open squares are (clockwise on the 25/12
versus 60/25 colour plot) 0, 1.7, 6.6, 10, 17, 26, 32, 37, 45, 57 and
72Myrs. The solid line shows the colours of a $\tau=20Myrs$ starburst,
calculated using equation (12), at the same ages. The symbol D denotes
the position of the 'Disc' component of RRC. To obtain the model
colours the model SEDs were convolved with the IRAS filter response
functions as explained in the IRAS Explanatory Supplement.}
\label{colours}  
\end{figure*}

 To synthesize the spectral energy distribution from a burst of star
 formation from those of individual GMCs, let us assume that at time
 $t$ after the onset of the starburst the star formation rate (or in
 our case the number of GMCs forming stars instantaneously with
 efficiency $\eta $) is $\dot{M}_*(t)$. If we further assume that the
 ensemble of clouds is optically thin, i.e. they don't shadow each
 other, then the flux from the burst is given by

\begin{equation}
F_\nu (t) = \int_0^t \dot{M}_*(t') S_\nu (t-t') dt' \label{eq12}
\end{equation}                 
where $S_\nu (t-t')$ is the flux from a GMC $t-t'$ years
after the onset of star formation at its centre.

 A useful parameterization for the star formation rate in a starburst
 which has been extensively used (e.g. Rieke et al 1980, Genzel et al 1998) is
 that of exponential decay, $\dot{M}_*(t') \sim e^{-t/\tau} $, where
 $\tau $ is some time constant. In the limit of large $\tau $ this
 approximates a scenario of constant star formation history. Under
 this assumption eqn (12) reduces to

\begin{equation}
 F_\nu (t) = \dot{M}_*(0) e^{-t/\tau} \int_0^t e^{t'/\tau} S_\nu (t') dt'
                           \label{eq13}
\end{equation}

In general, some stellar light will escape the GMC without being
absorbed by the dust shell, a situation we can't address exactly in
our present spherically symmetric model. To account for this,
{\em but only in the models fitted to M82 and NGC6090},  $S_\nu$
is corrected for this effect by allowing a fraction $f(t)$ of the
starlight to leak out. The emission from the dust is correspondingly
reduced to maintain flux conservation.  The leaking starlight may
suffer further absorption with the radiation reprocessed to infrared
radiation as well. Clearly there are a number of possibilities
here. The starlight may be absorbed locally (i.e. the GMC has some
kind of anisotropic density distribution, or optically thin holes), or
alternatively the starlight escapes the GMC and is absorbed by another
GMC or the diffuse medium (KS). In the first case the best
approximation would probably be to assume that the starlight is
reprocessed to the infrared with the same SED as that of the GMC
itself. In the second case it would be better to assume that the
starlight is reprocessed according to the SED of the more numerous
GMCs with ages of about 40-100Myrs.  In this paper, in particular in
the models fitted to M82 and NGC6090 in section 6, we consider only
the first case.  More details on the approximation used in those fits
can be found in Appendix A.

\section{Comparison with IRAS  data}

The IRAS data on galaxies have for the last decade or so provided the
yardstick by which theoretical models of their infrared emission
should be measured. While more extensive datasets are increasingly
becoming available with ISO and other projects (e.g. SCUBA), which
enable more detailed comparisons (and we illustrate this in the next
section for two well studied objects), it is still instructive to
compare our models with the IRAS colours.

 RRC proposed that the IRAS colours of galaxies can be understood in
 terms of 3 components: a `disc' component with infrared properties
 similar to those of cirrus clouds in our Galaxy, a starburst component
 with properties similar to those of compact galactic HII regions and
 an AGN component, which is associated with the dusty torus that is
 now an integral part of the standard AGN model. The latter component
 is not discussed further in this paper but instead we concentrate
 on the colours of starburst and Disc galaxies.

\begin{figure*}
\epsfig{file=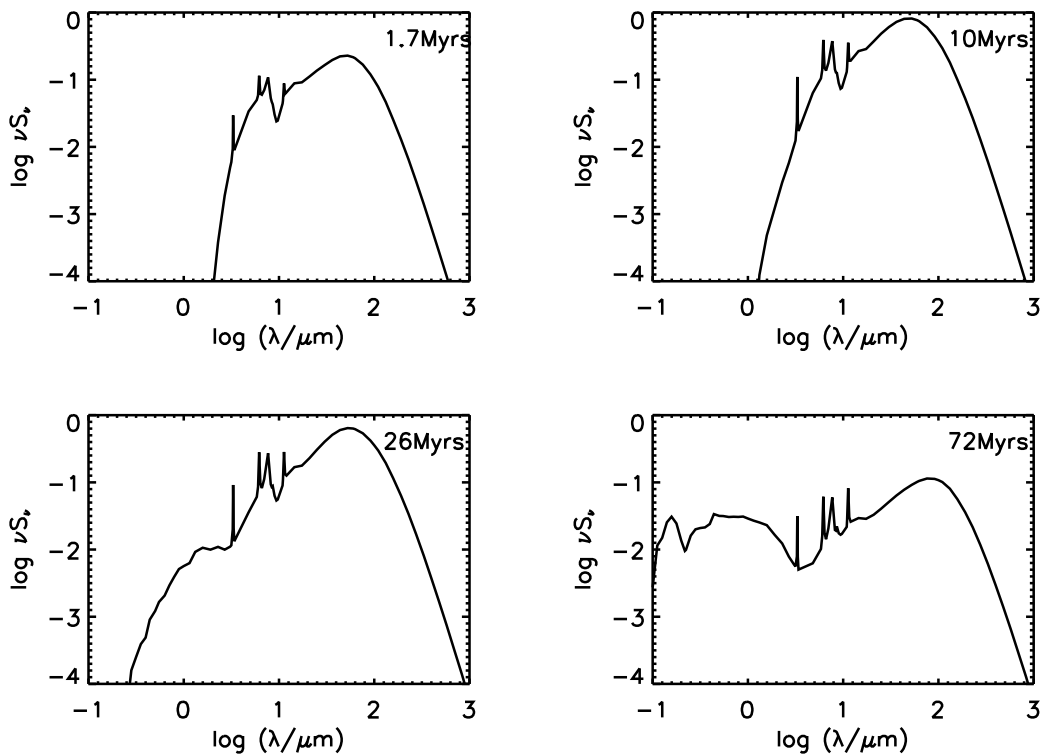, angle=0, width=15cm}
\caption{Spectral energy distributions of a $\tau = 20Myrs$ starburst
at different ages. The assumed GMC parameters are $M_{GMC}=10^7
 M_{\odot}, \eta=0.25, n_{av}=300 cm^{-3}$. The vertical scale is
 arbitrary but constant for the four models. Note the bolometric
 luminosity of the starburst peaks at about 10-20Myrs.}
\label{starbursts}  
\end{figure*}

In Figure 2 we plot the IRAS colours of the starburst galaxies in the
sample of RRC (selected to have good quality fluxes in all four IRAS
bands) and indicate the position of their disc component (D). Also
indicated in Figure 2 are the positions of M82 and NGC6090 on the
colour-colour diagrams. Disc galaxies cover the part of the
colour-colour diagram between the starbursts and D, although there is
some overlap with the starburst galaxies. RRC proposed that the overlap
is due to mixing of the starburst and Disc components. Rowan-Robinson
(1992) also showed that some of the variation in the colours of the
disc galaxies may be due to the intensity of the radiation field.

If we consider first the predicted colours of the sequence of GMCs
(dotted line), we see that they span the entire range of observed
galaxy colours. This suggests that a weighted sum of emission from
such a family of GMCs (which is what equation (12) essentially is) may
explain the observed galaxy colours. To test this we have computed the
colours of a galaxy that experienced an exponentially decaying burst
with $\tau=20$Myrs (solid line). The predicted colours in the age
range 0-72Myrs nicely match the spread in the colours of starburst
galaxies. Furthermore, they predict a correlation of the 100/60 ratio,
and an anticorrelation of the 25/12 one, with age. The predicted mean
age of starburst galaxies ($\sim 50$Myrs) agrees remarkably well with
other estimates from Br${\gamma}$ equivalent widths and CO indices
(Goldader et al 1997) as well as ISO spectroscopy (Genzel et al
1998). The predicted colours are not very sensitive on the assumed
value of $\tau $.  The colours for the constant star formation case, a
scenario more appropriate for disc galaxies, also lie on the same
track but are packed towards the M82 end. This may have significant
implications for the origin of the far-IR luminosity of disc galaxies
as we discuss further in section 8.  The conclusion from this analysis
is therefore that the age of the burst can account for some of the
variation in IRAS galaxy colours attributed by RRC to mixing with a cirrus
component.

 There are two more features of the colours of the sequence of GMCs
 that are worth mentioning. First the discontinuity at $t=20Myrs$ is
 due to the switch from the HII to supernova phase.  Clearly, how
 quickly (and how far) the tracks move to the right of the
 colour-colour diagrams will depend on the details of the supernova
 phase which we treat very crudely in this paper. If the supernova
 ejecta expel the dust further away than we have assumed, then (in the
 absence of any other heating source) we would expect the dust to cool
 down further and the 60/25 and 100/60 ratios to increase.
 The discontinuity in the colours may also be an artifact of our
 simplified treatment of the GMC evolution. In particular, stellar
 winds from young stars may impart as much kinetic energy in the ISM
 as the type II supernovae. 
  Secondly, the apparent `retrograde' motion of the tracks, best seen in the
 25/12 versus 60/25 colour-colour diagram, is due to the radiative
 transfer effect, highlighted in section 3, which leads to the
 temporary heating up of the dust shell despite the fact that the
 radiation field is diminishing. The reality of these features can
 only be assessed by more detailed simulations of the evolution of
 GMCs.

 In Figure 3 we give the SEDs of the starburst discussed above
 ($\tau=20$Myrs) for some representative ages. No correction for
 leaking starlight is applied to the models. A number of features are
 worth highlighting: (1) there is a general tendency for the peak of
 the SED to shift to longer wavelengths with age. (2) The PAH features
 get stronger with age. (3) the 9.7$\mu m$ silicate feature gets
 shallower with age. (4) while the optical/UV light of the youngest
 bursts is almost completely obscured, the oldest starbursts are
 predicted to emit significantly in the optical/UV. This illustrates
 the extent to which old diffuse (and therefore optically thin) GMCs
 dominate the emission of old bursts.

\section{A model for M82}

\subsection{Observational constraints on M82}

M82 being the nearest example of a starburst  (D=3.25Mpc Tammann
\& Sandage 1968) has received a lot of attention both observationally and
theoretically.  The luminosity of the starburst ($3\times
10^{10}L_\odot$ Telesco \& Harper 1980) is sufficient to outshine any
pre-existing stellar population in the nucleus of this dwarf
galaxy. The IR luminosity arises mostly in the central 200x400pc
region (13'' x 26'') which shows a bilobal distribution and is aligned
with a central stellar bar. The bar is mainly made up of an old
stellar population (Telesco et al 1991) and it is thought that it has
probably formed about $10^8$ years ago during the interaction of M81
with M82.  Radio interferometry suggests a supernova frequency of 1
every 10-20 years in the central region which implies a high star
formation rate (Muxlow et al 1994).

 Rieke et al (1980), in their classic study of M82, drew attention to
the fact that the starburst emission is spatially correlated at 10-,
20$\mu m$ and radio wavelengths. The emission region constitutes an
elongated structure aligned with the major axis of the galaxy.  The
broad correspondence of the mid-IR emission with the radio was also
confirmed by Telesco \& Gezari (1992). The map of Telesco \& Gezari
shows the two prominent peaks seen in earlier observations but their
higher spatial resolution also allowed them to resolve a number of
clumps with characteristic radii of about 20pc. This provides further
confirmation of the basic assumption of our model that the starburst
is made of an ensemble of massive GMCs.

 Rieke et al also pointed out that the starburst shows a completely
different appearance in the K band with the `nucleus' of the galaxy
being the prominent feature. There is no 10$\mu m$ feature associated
with the nucleus. In the K band there is also extended emission along a
stellar bar about a kpc in extent.  Rieke et al also found that 2$\mu
m$ spectra of the nucleus show the absorption bands characteristic of
giants and supergiants.  Their interpretation of these findings is
that the nucleus represents the first generation of stars formed about
$5 \times 10^7$ yrs ago in an (exponentially decaying) starburst which
has since propagated outwards.

Shen \& Lo (1995) mapped the CO emission from M82 with 2.5''
resolution. Their maps revealed unresolved ($<30pc$) structure in the
CO emission. They suggest that most of the CO emitting gas (and
therefore the starburst) is located in molecular spilar arms at 125
and 390pc from the nucleus. They also resolved the previously known
double-peaked structure in several peaks.

The task of utilizing the multiwavelength data for M82 is complicated
by the fact that the good spatial and spectral resolution data at the
shorter wavelengths $\lambda \leq 60 \mu m$ can not be exactly matched
by the far-IR and submm data.  A scan by Telesco \& Harper (1980) at
58$\mu m$ (where the spectral energy distribution peaks) shows that
the intrinsic source size is about 30'' very similar to that at 10$\mu
m$.  Hughes et al (1994) deconvolved the source size of M82 at various
wavelengths. They find that between 10-40$\mu m$ the source size is
fairly constant (25x8 arcsec$^2$) but between 100$\mu m$ and 1.3mm it
is slightly larger (37x10 arcsec$^2$).

 While an IRAS LRS spectrum (with an aperture large enough to include
 the whole galaxy) is available for M82, because of the importance of
 the mid {\em and} near-IR wavelength range in constraining the model,
 we use instead the 2-13$\mu m$ spectrophotometry by Willner et al
 (1977) with a 30'' aperture.  The 10$\mu m$ spectrum looks similar to
 that of Roche et al (1991) and Jones \& Rodriguez-Espinosa (1984) in a number
 of positions with 3'' which lends further support to the idea that
 what we see in the central 30'' is an ensemble of similar
 star-forming regions.  The IRAS LRS spectrum also shows a similar
 spectrum although the 10$\mu m$ feature appears a little shallower.
 The flux in the IRAS LRS spectrum is also about a factor of 2 higher
 than the spectrum of Willner but also about 50\% higher than the
 broad band 12$\mu m$ point. The 10$\mu m$ growth curves of Rieke et
 al (1980) also suggest that the total mid-IR flux may be higher than
 that in the 30'' aperture by about 50\%.  The K band photometry of Rieke
 in a 30'' aperture matches that of Willner et al in a similar aperture
 rather well. The K flux in the 30'' aperture is about 30\% of the
 total (Rieke et al, Ichikawa et al 1995).

\begin{figure}
\epsfig{file=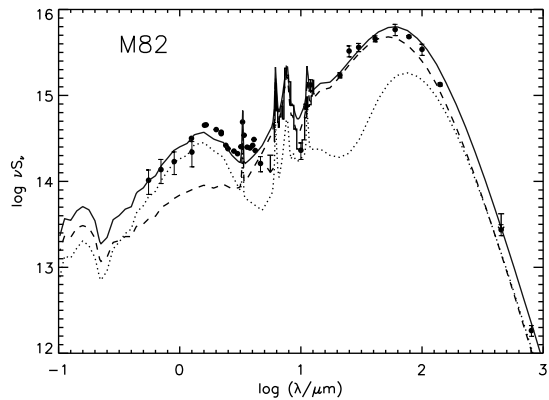, angle=0, width=8cm}
\caption{Fit to the spectrum of M82 with the starburst model. As
 discussed in the text the model assumes that M82 has experienced two
 exponentially decaying bursts over the last 26.5Myrs. The first one
 ($26.5Myrs, \tau =2Myrs$) is shown by the dotted line and the second
 one ($16.5Myrs, \tau =6Myrs$) by a dashed line. The total emission is
 given by the solid line. Data from Willner et al 1977, Johnson et al
 (1966), RRE and references therein}\label{M82}
\end{figure}

 In conclusion, the data plotted in Figure \ref{M82} are probably a
 good description of the spectrum of the starburst in M82 but it
 should be borne in mind that the far-IR and submm data may include
 contributions from a region more extended than the central starburst.
 Having experimented with a range of combinations of age and decay
 time $\tau $ we could not obtain a fit to the multi-wavelength data
 with a single burst scenario. The fundamental problem with a single
 burst (in the context of our model) is that M82 shows characteristics
 for both a young burst (warm SED and fairly deep 10$\mu m$
 absorption) and an older burst (strong 1-5$\mu m$ continuum).

 Our model fit, shown in Figure \ref{M82},
assumes that M82 experienced two bursts of star formation. The first
one occured 26.5Myrs ago and had a very steep exponential decay ($\tau
=2Myrs$). This burst is now responsible for most of the K band light
and about half of the far-IR and submillimeter emission. The second
burst occured 16.5Myrs ago and decayed more slowly ($\tau
=6Myrs$). The GMCs in both bursts are assumed to leak 20$\%$ of their
starlight after 10Myrs which subsequently suffers a visual extinction
of 1.5 magnitudes. The two bursts are predicted to contribute roughly
equal amounts of UV flux. The spectrophotometry of Willner et al shows
an excess over the model longwards of the 3.3$\mu m$ feature (possibly
associated with a quasi-continuum similar to that underlying the
longer wavelength features) which we cannot match with our present
grain model. Note that while the 10$\mu m$ spectrum of the younger
burst provides an excellent fit to the shape of the observed spectrum,
the addition of the older burst with its flatter spectrum dilutes the
absorption feature. Our model could be in better agreement with the
10$\mu m$ feature and the extended far-IR to submm emission if at
least some of the dust shells associated with the older burst have
expanded beyond the central 30 arcsec. In fact there is evidence
from the map of Hughes et al (1994) that some submm emission is
associated with the outflows along the minor axis of the galaxy.

\section{A model for NGC6090}

 Acosta-Pulido et al (1996) presented ISO spectrophotometry and
extensive multi-wavelength photometry for NGC6090. This galaxy is
about 10 times more luminous than M82 and its optical image shows a
disturbed morphology and signs of a recent interaction. UV to K band
data for this galaxy are also available (Gordon et al 1997). The IRAS
colours of NGC6090 show evidence for cold dust. RRC attributed that to
a significant contribution from cirrus in this galaxy. It is therefore
not surprising that the emission from an M82 type starburst has to be
supplemented by colder dust to fit the data for $\lambda > 120 \mu m$
(Acosta-Pulido et al).

\begin{figure}
\epsfig{file=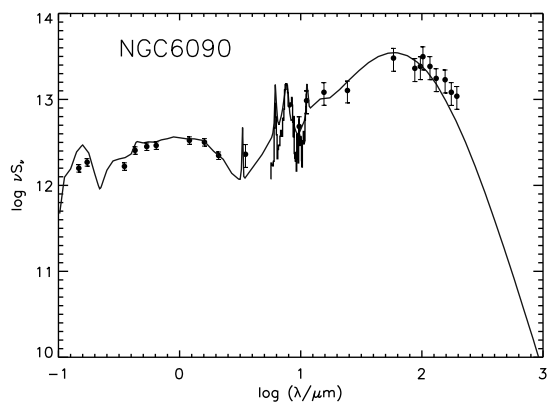, angle=0, width=8cm}
\caption{Fit to the spectrum of NGC6090 with an exponentially decaying
 starburst model ($t=64Myrs, \tau=50Myrs$).
 Data from Acosta-Pulido, J.A., {\em et al}, (1996), and Gordon et al
 (1997)}\label{ngc6090}   
\end{figure}

 As it is clear from Figure 2, the IRAS colours of NGC6090 predict
 an `old' burst scenario.
 In Figure \ref{ngc6090} we test this by comparing such a model to the
 multiwavelength data. We find that a good fit can be obtained with
 $t=64$Myrs, $\tau=50$Myrs. As in the M82 model we have assumed that
 after 10Myrs 20$\% $ of the starlight from each GMC escapes (and
 subsequently suffers 1.5mag of visual extinction), but we find that
 adequate fits to the data can be obtained even if we have no leakage.

 The remaining discrepancy between the model and the far-IR
 observations suggests that some emission from even colder dust in the
 starburst or general interstellar medium of this galaxy (cirrus) may
 be contributing at the longer wavelengths.  Colder dust in the
 starburst could for example arise if the supernova explosions are
 more effective (as we discuss in section 5) in destroying the GMCs
 and expelling the dust away from the heating sources. Evidence of
 very cold dust ($T \sim 10K$) has been found in some normal galaxies
 by recent ISO observations (e.g. Kr\"ugel et al 1998). The same
 authors, however, find no evidence for cold dust in the starburst
 galaxies in their sample. It will be interesting to assess how common
 is this far-IR excess in starburst galaxies as it will provide strong
 constraints to theoretical models.

\section{Summary and Discussion}

We have described an evolutionary scheme for massive GMCs centrally
illuminated by a stellar cluster. The defining characteristic of this
scheme is that the HII regions formed by the ionizing stellar
radiation compress the neutral gas and dust in the GMC to a narrow
shell. This naturally explains why the maximum temperature of the
large grains in HII regions appears to be lower than the sublimation
temperature of the graphite and silicate grains. It also explains why
the near to mid-IR spectrum of HII regions and starburst galaxies is
dominated by the PAH emission. These trends are consistent with the
findings of Helou et al (1998) from ISO mid-IR spectrophotometry. The
evolution of the GMC after about $10^7$yrs is more uncertain but we
show that even a modest rate of expansion at the interstellar velocity
dispersion is sufficient to lead to the formation of a diffuse cold
shell by about $10^8$yrs. The infrared spectrum of these shells is
very similar to infrared cirrus clouds. We associate these shells with
the HI superstructures observed in our galaxy and other galaxies
(Heiles 1979) and usually associated with large star clusters.

The sequence of GMC spectral energy distributions we have computed
cover the range of observed galaxy SEDs as observed by IRAS. This
leads us to believe that the observed spectra of galaxies could be
modelled in terms of this evolutionary scheme and constrain their star
formation history. Exponentially decaying bursts are shown to account
satisfactorily for the colours of starburst galaxies in the sample of
RRC.  The age of the burst is shown to be an important contributing
factor to the spread in the galaxy colours and can mimick (to some
extent) the effect of mixing with a cirrus component. Our application
of these models to M82 and NGC6090 has shown that good fits to the UV
to mm data on these galaxies can be obtained with the models.  In M82
we find evidence for two bursts separated by 10Myrs. Our model is
similar in this respect to that of Rieke et al (1993) who find that
two Gaussian bursts (separated by 8-25Myrs) can account for a range of
observational constraints. Theoretical support for this periodic burst
scenario is also provided by the models of Kr\"ugel \& Tutukov (1993).
NGC6090 is best fitted by an older burst (64Myrs). There is evidence
for a colder dust component in this galaxy. It is not clear whether
that is due to cirrus or colder dust in the starburst.

The colours predicted for a continuous star formation scenario, one
more appropriate for disc galaxies, falls short of explaining the
colours of disc galaxies. Although a more thorough exploration of the
parameter space is needed before any definite conclusions can be
drawn, this result is not entirely surprising for the following
reason. In starburst galaxies, almost by definition, the emission from
the recently formed stars outshines the old stellar population so the
effect of the latter on our models is probably not very
significant. In disc galaxies, however, where the star formation rate
is lower, the old stellar population is more important and needs to be
taken into account when considering the energy balance and emission of
GMCs, especially the older ones. The effect of this would be,
neglecting any change in the spectral shape, to boost the luminosity
of the old GMCs and therefore their contribution to the overall galaxy
emission. This provides an indirect argument in favour of the idea
that old stars make a significant contribution to the far-IR
luminosity of disc galaxies (Waltebros \& Greenwalt 1996).

The wealth of data on M82 allows us to apply further checks on the
validity of this model.  O'Connell et al (1995) imaged the central few
hundred parsecs of M82 with the HST in the V and I bands and discovered
a complex of over 100 luminous star clusters. The clusters do not seem
to be associated with X-rays, infrared or radio compact sources but
instead with the less obscured regions in the galaxy such as the
`nucleus' and the periphery of the dust lane. The size of these
clusters is estimated to be about 3pc, consistent with being the
remnants of the GMCs in our model for the starburst.  The fact that
M82 is viewed almost edge-on complicates somewhat the interpretation
of these observations but the (at least) partial segregation of the
optical and infrared emitting regions suggests that they arise from
stellar populations at different evolutionary stages, as in our
model. This also suggests that the assumption of spherical symmetry is
reasonable at least for the young ($t < 10$Myrs) GMCs.

Satyapal et al (1997) studied the stellar clusters in the central
500pc of M82 with near-IR spectroscopy and high resolution
imaging. Their findings support a picture in which the typical age of
the stellar clusters is $10^7$ yrs. They also find a correlation
between the age of the clusters and distance from the centre
suggesting that the starburst is propagating outwards at a speed of
about 50km/s. This picture is also supported by spectro-imaging of M82
at 3.3$\mu m$ which shows evidence for dissociation of PAH molecules
in the `nucleus' (Normand et al 1995).  McLeod et al (1993) also found
that the 3.3$\mu m$ feature is about twice as strong at a position
offset by about 8'' from the nucleus.  The centre of M82 has been
known for some time (Axon \& Taylor 1978) to show evidence for a
biconical outflow along its minor axis. This is now widely believed to
be due to a galactic wind, a general characteristic of starburst
galaxies, powered by stellar winds and supernovae. Numerical
hydrodynamic simulations (Suchkov et al 1994) suggest that such
outflows can develop under quite general conditions at the center of
starbursts but their morphology depends very much on the energy
deposition rate and density of the medium. It is likely that the
galactic wind was powered by the earlier burst and that the
interaction of the wind with gas in the plane of the galaxy triggered
or at least contributed to the outwardly propagating secondary burst.

 Telesco \& Gezari (1992) noted that there is no correspondence
 between the peaks of the radio emission in M82 (presumed to be young
 supernova remnants) and those at 12.4$\mu m$. This suggests that the
 mid-IR peaks in M82 are associated with GMCs in which supernova
 activity has not yet started or in which the supernova rate is still
 very low. This observation is in excellent agreement with our M82
 model in which the mid-IR emission is predominantly due to the
 younger burst.

\begin{figure}
\epsfig{file=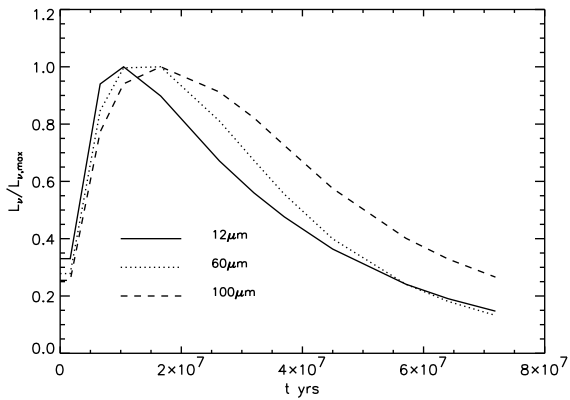, angle=0, width=8cm}
\caption{
Predicted evolution of the luminosity (normalized to its peak) of a
starburst with $\tau =20Myrs$ at different infrared wavelengths.}
\end{figure}

 As is evident from Figure 3 the luminosity of a starburst galaxy is
 predicted to change quite considerably with age. We illustrate this
 more clearly in Figure 6 where we plot the luminosity at different IRAS
 wavelengths (normalized to its peak) as a function of age for the
 same model parameters used in Figure 3 ($\tau =20Myrs$). The
 luminosity rises sharply in the first 10Myrs to peak at an age which
 depends on the wavelength. It then declines sharply by up to a factor
 of 5 with age. We have omitted the 25$\mu m$ curve as it is very
 similar to that at 60$\mu m$. There are a number of interesting
 implications arising from this result for the detectability of
 starburst galaxies. It implies for example that
 mid-IR surveys would preferentially detect younger starbursts than
 far-IR surveys. It also implies that near the detection limit of such
 surveys up to half the  galaxies that experienced a burst in the last
 $7 \times 10^7$ years may be missed because they are
 either too young or too old. This may have implications for estimates
 of the star formation rate over volumes of space as it introduces
 another form of bias.

 In conclusion, the evolutionary scheme we have put forward in this
 paper seems to be in very good agreement with an array of
 observational evidence on M82 and is consistent with the infrared
 properties of other starburst galaxies. It promises to be useful for
 the interpretation of the growing datasets on infrared galaxies and
 we plan such studies in future work. The scheme also lends itself for
 incorporation into  simulations of galaxy formation and evolution and
 this was one of the prime motivations for its development.

 While useful constraints on the star formation history and stellar
 populations of galaxies can be obtained from this model, it will be
 of much interest to explore in the future the effect of deviations
 from spherical symmetry especially at the later stages of GMC
 evolution. Hydrodynamical simulations of the evolution of
 star-forming molecular clouds which take into account the effects of
 ionization, stellar winds and multi-supernova explosions in dense
 environments should also take high priority. Only then will we be
 able to take full advantage of the data ISO, SCUBA, WIRE, SIRTF,
 NGST, VLT, FIRST, PLANCK, SOFIA, ALMA etc. are
 expected to yield over the next decade or so.

\section*{Acknowledgments}

 AE acknowledges support by PPARC. This work has made use of the NASA
 Extragalactic Database (NED). We thank an anonymous referee for
 useful comments and suggestions.

\section*{Conclusions}

\section*{APPENDIX: Corrections for non-spherical geometry}

Let the monochromatic luminosity of a spherical GMC of optical depth 
$\tau_V$ (centrally illuminated by a stellar cluster) at frequency $\nu$
be $4\pi S_\nu^s$ whereas that of the stellar cluster (in the absence
of any dust) be  $4\pi S_\nu^*$. By definition, $\int_o^\infty S_\nu^s
 d\nu \equiv \int_o^\infty S_\nu^* d\nu$. If the optical depth of the GMC is
$\tau_V^l (< \tau_V)$ for a fraction $f$ of the sky centred on the
 star cluster,  then the luminosity of such a non-spherical GMC can be
 approximated by,

$$
L_\nu= 4 \pi [(1-f) S_\nu^s + f e^{-{{C_{e,\nu}} \over  {C_{e, V}}}
\tau_V^l} S_\nu^* + f {
{ \int_o^\infty (1 - e^{-{{C_{e,\nu} \over {C_{e,V}}}}})  S_\nu^*
 d\nu}
 \over
{ \int_o^\infty  S_\nu^*  d\nu}} S_\nu^s]$$
where $C_{e,\nu}$ is the dust extinction cross-section at frequency
$\nu$.

 The first term assumes that the GMC emits isotropically with the same
 spectral energy distribution as the spherical GMC. The factor $1-f$,
 however, accounts for the fact that a fraction $f$ of the sky is
 covered by a lower (or even zero) optical depth. The remaining two
 terms attempt to correct for this complication. The second term
 gives the stelar light that is transmitted through the optically thin
 holes. The third term assumes that the light absorbed by dust in the
 optically thin holes is reprocessed to the infrared with the same
 spectral energy distribution as that of the spherical GMC.

 If a large number of non-spherical, but otherwise identical, GMCs
 viewed at different orientations are present in a starburst, then in
 equations 12 and 13 one can use their average emission $S_\nu \equiv
 {{L_\nu} \over {4\pi}}$. While this approximation conserves energy and
 is probably close to the best one can do for correcting for
 non-spherical geometry, without doing the actual calculation, its
 usefulness is bound to be limited as some of the assumptions
 (e.g. isotropic emission from a non-spherical optically thick system)
 are known to be false (e.g. Efstathiou \& Rowan-Robinson 1995).


\begin{thebibliography}{99}


\bibitem{} Acosta-Pulido, J.A., {\it et al.}, 1996, AA, 315, 121.

\bibitem{} Allamandola, L.J., Tielens, A.G.G.M., Barker, J.R., 1989,
  ApJS, 71, 733.

\bibitem{}  Axon, D.J., \& Taylor, K., 1978, Nature, 274, 37.

\bibitem{} Bruzual, A \& Charlot, S., GISSEL user guide, unpublished.

\bibitem{} Churchwell, E., Wolfire, M.G., Wood, D.O.S., 1990, ApJ, 354, 247.

\bibitem{} Colbert, J.W., et al, 1998, astro-ph/9810188.

\bibitem{} Dame, , T.M., Elmegreen, B.G., Cohen, R.S., \& Thaddeus,
P., ApJ, 305, 892.

\bibitem{} Desert, F.-X., Boulanger, F., Puget, J.L., 1990, AA, 237, 215.

\bibitem{} Downes, D., Solomon, P.M., 1998, ApJ, 507, 615.

\bibitem{} Draine, B.T., Lee, H.M., 1984, ApJ, 285, 89.

\bibitem{} Draine, B.T., 1989, ApJ, IAU Symposium 135, 313. 

\bibitem{} Edoh, O., 1983, Ph.D. thesis, Univ. of Arizona.

\bibitem{} Efstathiou, A., Rowan-Robinson, M., 1990, MN, 245, 275. 

\bibitem{} Efstathiou, A., Rowan-Robinson, M., 1994, MN, 266, 212. 

\bibitem{} Efstathiou, A., Rowan-Robinson, M., 1995, MN, 273, 649. 

\bibitem{} Efstathiou, A., Siebenmorgen, R., 1999, in preparation. 

\bibitem{} Elmegreen, B.G.,  1985,  in Protostar and Planets II, p.33-58.

\bibitem{} Franco, J., Tenorio-Tagle, G., Bodenheimer, P., 1990, ApJ,
349, 126.

\bibitem{} Genzel, R.,  et al 1998, ApJ, 498, 579.

\bibitem{} Goldader, J.D., Joseph, R.D., Doyon, R., Sanders, D.B.,
1997, ApJ, 474, 104.

\bibitem{} Gordon, K.D., Calzetti, D., \& Witt, A.N., 1997, ApJ, 487, 625.

\bibitem{} Guhathakurta, P., Draine, B.T., 1989, ApJ, 345, 230.

\bibitem{} G\"{u}sten, R., 1989, in IAU Symposium 136, The Center of the
Galaxy, ed.M.Morris, Dordrecht(Reidel), p.89.

\bibitem{} Heiles, C., 1979, ApJ, 229, 533.

\bibitem{} Helou, G., et al, 1998, In `Astrophysics with Infrared
Surveys: A prelude to SIRTF', ed.M.Bicay, in press.

\bibitem{} Hughes, D.H., Gear, W.K., \& Robson, E.I., 1994, MNRAS, 270, 641.

\bibitem{} Hughes, D., et al 1998, Nature, 394, 241.

\bibitem{} Ichikawa, T., Yanagisawa, K., Itoh, N., Tarusawa, K., van
Driel, W., Ueno, M. 1995, AJ, 109, 2038.

\bibitem{} Johnson, H.L., 1966, ApJ, 143, 187.

\bibitem{} Jones, B., \& Rodriguez-Espinosa, J.M., 1984, ApJ, 285, 580.

\bibitem{} Kennicutt, R.C., 1998, ApJ, 498, 541.

\bibitem{} Kr\"ugel, E.,  \& Walmsley, C.M., 1984, AA, 130, 5. 

\bibitem{} Kr\"ugel, E., Siebenmorgen, R., 1994, AA, 282, 407. (KS)

\bibitem{} Kr\"ugel, E., Tutukov, A.V., 1993, AA, 275, 416. 

\bibitem{} Kr\"ugel, E., Siebenmorgen R., Zota, V., Chini, R., 1998,
AA, 331, L9.

\bibitem{} Larson, R.B., 1988, In 'Galactic and Extragalactic star formation',
 eds. R.E.Pudritz, M.Fich. Dordrecht:Kluwer.

\bibitem{} Lilly, S.J., Le Fevre, O., Hammer, F., \& Crampton, D.,
1996, ApJ, 460, L1.

\bibitem{} Madau, P., Ferguson, H.C., Dickinson, M.E., Giavalisco, M.,
Steidel, C.C., Fruchter, A., 1996, MNRAS, 283, 1388.

\bibitem{} Mathis, J.S., 1990, ARAA, 28, 37.

\bibitem{} McCray, R.,  \& Kafatos, M., 1987, ApJ, 317, 190.

\bibitem{} McCleod, K.K., Rieke, G.H., Rieke, M.J., Kelly, D.M., 1993,
ApJ, 412, 111.

\bibitem{} Meurer, G.R., Heckman, T.M., Leitherer, C., Kinney, A.,
Robert, C., Garnett, D.R., 1995, AJ, 110, 2665.

\bibitem{} Muxlow, T.W.B., Pedlar, A., Wilkinson, P.N., Axon, D.J.,
Sanders, E.M., De Bruyn, A.G., 1994, MNRAS, 266, 455.

\bibitem{} Normand, P., Rouan, D., Lacombe, F., Tiphene, D., 1995, AA,
297, 311

\bibitem{} O'Connell, R.W., Gallagher III, J.S., Hunter, D.A., Colley,
W.N., 1995, ApJ, 446, L1.

\bibitem{} Pearson, C., Rowan-Robinson, M., 1996, MN 283, 174

\bibitem{} Puget, J.L., Leger, A., 1989, ARAA, 27, 161.

\bibitem{} Rieke, G.H., Lebofsky, M.J., Thompson, R.I., Low, F.J.,
Tokunaga, A.T., 1980, ApJ, 238, 24.

\bibitem{} Roche, P.F., Aitken, D.K., Smith, C.H., Ward, M.J., 1991,
MN, 248, 606. 

\bibitem{} Rowan-Robinson, M., 1980, ApJS, 44, 403.

\bibitem{} Rowan-Robinson, M., 1992, MN 258, 787

\bibitem{} Rowan-Robinson, M., Crawford, J., 1989, MN 238, 523. (RRC)

\bibitem{} Rowan-Robinson, M., Efstathiou, A., 1993, MN 263, 675. (RRE)

\bibitem{} Rowan-Robinson, M., et al 1997, MN, 289, 490.

\bibitem{} Satyapal, S., Watson, D.M., Pipher, J.L., Forrest, W.J.,
Greenhouse, M.A., Smith, H.A., Fischer, J., Woodward, C.E., 1997, ApJ,
483, 148.

\bibitem{}  Savage, B.D., \& Mathis, J.S., 1979, ARAA, 17, 73.

\bibitem{} Shen \& Lo, 1995, ApJ, 445, L99.

\bibitem{} Spitzer, L., 1978, `Physical Processes in the interstellar
medium', Wiley-Interscience.

\bibitem{} Suchkov, A.A., Balsara, D.S., Heckman, T.M., Leitherer, C.,
1994, ApJ, 430, 511.

\bibitem{}  Siebenmorgen, R., Kr\"ugel, E., 1992, AA, 259, 614. (SK92) 

\bibitem{}  Siebenmorgen, R., Kr\"ugel, E., Mathis, J.S., 1992, AA,
266, 501.  

\bibitem{} Siebenmorgen, R., 1993, ApJ, 408, 218.

\bibitem{} Steidel, C.C., Hamilton, D., 1992, AJ, 104, 104.

\bibitem{} Solomon, P.M., Rivolo, A.R., Barrett, J., \& Yahil, A.,
1987, ApJ, 322, 101. 

\bibitem{} Tammann, G.A., \& Sandage, A., 1968, ApJ, 151, 825.

\bibitem{} Telesco, C.M., \& Harper, D.A., 1980, ApJ, 235, 392.

\bibitem{} Telesco, C.M., Joy, M., Dietz, K., Decher, R., Campins, H.,
1991, ApJ, 369, 135.

\bibitem{} Telesco, C.M., \& Gezari, D.Y., 1992, ApJ, 395, 461.

\bibitem{} Tenorio-Tagle, G., \& Bodenheimer, P., 1988, ARAA, 26,145.

\bibitem{} Waltebros, R.M., \& Greenwalt, B., 1996, ApJ, 460, 696.

\bibitem{} Whitmore, B.C., \& Schweizer, F., 1995, AJ, 109, 960.

\bibitem{} Willner, S.P., Soifer, B.T., Russell, R.W., Joyce, R.R.,
Gillett, F.C., 1977, ApJ, 217, L121.


\end{thebibliography}
\end{document}